\documentclass[11pt]{article}
\usepackage{amsfonts}
\usepackage[a4paper,margin=2.5cm]{geometry}
\usepackage{amsmath,amssymb,amsthm}

\newtheorem{proposition}{Proposition}
\newtheorem{lemma}{Lemma}
\newtheorem{conjecture}{Conjecture}
\theoremstyle{definition}
\newtheorem{definition}{Definition}
\newtheorem{corollary}{Corollary}
\newtheorem{remark}{Remark}
\newcommand{\Nat}{\mathbb{N}}

\begin{document}

\title{More on Higher Order Busy Beaver Functions}
\author{Zining Cao \\
College of Computer Science and Technology\\
Nanjing University of Aeronautics and Astronautics, Nanjing 211106, China\\
\texttt{caozn@nuaa.edu.cn}}
\date{}
\maketitle

\begin{abstract}
In this paper, we extend the Busy Beaver function to a class of higher order Busy Beaver functions based on Turing oracle machine. We prove some results about the relation between decidability of number theoretical formula and higher order Busy Beaver functions, and the relations between computability of max-min partial recursive functions, limit partial recursive functions and higher order Busy Beaver functions. We also present some conjectures on higher order Busy Beaver functions.
\end{abstract}

\section{Introduction}

\label{sec:intro}

The Busy Beaver function $BB(n)$, introduced by Rad\'o \cite{rado}, maps $n$
to the maximal number of steps taken by any $n$-state Turing machine that
halts on the blank tape. It grows faster than any computable function and
marks a natural boundary of computability.

Aaronson \cite{aaronson} sketched a \emph{super Busy Beaver function}: the
step-maximization over $n$-state Turing machines equipped with an oracle for
$BB$ itself, and observed that such a function eventually dominates every
computable function even at slightly smaller arguments. However, a fully
formal definition of this idea --- in particular of the iteration to higher
orders and of the role of the input string --- was not carried out in \cite%
{aaronson}. The purpose of the present paper is to supply such a definition
and to develop its basic computability-theoretic consequences.

We emphasize the relation to known theory. It is classical (see, e.g., \cite%
{rogers,post}) that the $\Sigma_i$/$\Pi_i$ levels of the arithmetical
hierarchy correspond exactly to decidability by Turing machines with a $%
0^{(i)}$ (the $i$-th Turing jump) oracle, and that oracles for $BB$-type
functions compute the jumps. Our Proposition~\ref{prop:arith} is therefore a
reformulation of this classical correspondence in the language of Busy
Beaver oracles; its value lies in the fact that the oracle is \emph{defined
by a step-maximization over machines}, which makes the connection between
the hierarchy and machine-step counting explicit. The genuinely new elements
of this paper are: the parameterized definition $BB(s,m,n)$ with input
string $s$ (Definition~\ref{def:hobb}); the exact characterization of the
machine-computable-with-$BB$-oracle functions as the max-min partial
recursive functions, including a complete coding proof (Lemma~\ref{lem:deriv}%
); and the growth comparison of Proposition~\ref{prop:ordergrowth} together
with Conjectures~\ref{conj:n} and~\ref{conj:s} for which we give heuristic
evidence.

The paper is organized as follows. Section~\ref{sec:prelim} reviews Turing
machines and the classical Busy Beaver function. Section~\ref{sec:def} gives
the formal definition of higher order Busy Beaver functions. Section~\ref%
{sec:results} presents the decidability and computability results, with full
proofs of the coding lemmas. Section~\ref{sec:growth} presents growth
comparisons and conjectures with heuristics. Section~\ref{sec:conc}
concludes.

\section{Turing machines and the Busy Beaver function}

\label{sec:prelim}

We consider deterministic Turing machines over the alphabet $\{0,1\}$, with
a two-way infinite tape, a finite set $Q$ of states with initial state $q_0$%
, and a transition function which, upon reading the current tape symbol,
either (i) writes a symbol, moves the head one square left or right, and
enters a next state, or (ii) halts. This is the standard model of \cite%
{sipser}.

\begin{definition}
\label{def:tm} A deterministic Turing machine is a tuple $M=\langle
Q,\Gamma,\delta,q_0\rangle$ where

\begin{enumerate}
\item $Q$ is a finite, non-empty set of states;

\item $\Gamma=\{0,1\}$, where $0$ is the blank symbol;

\item $\delta :Q\times \Gamma \rightharpoonup Q\times \Gamma \times \{L,R\}$
is a partial transition function; if $\delta $ is undefined on the current
pair (state, symbol), the machine halts;

\item $q_{0}\in Q$ is the initial state.
\end{enumerate}
\end{definition}

For a machine $M$, let $\mathit{step}(M)$ be the number of steps $M$ takes
before halting on the all-$0$ initial tape (including the final halt step),
and $\mathit{step}(M)=\infty$ if $M$ never halts. Let $T(n)$ be the set of $%
n $-state machines.

\begin{definition}
The Busy Beaver function is
\begin{equation*}
BB(n)\;=\;\max\{\,\mathit{step}(M) \;:\; M\in T(n),\ \mathit{step}%
(M)<\infty\,\}.
\end{equation*}
\end{definition}

\section{Higher order Busy Beaver functions}

\label{sec:def}

\subsection{Turing machines with a fixed oracle sequence}

\begin{definition}
\label{def:oracle} A \emph{deterministic $m$-order oracle Turing machine}
(with respect to a fixed sequence of total functions $F_1,\dots,F_m$, $%
F_i:\{0,1\}^{*}\rightharpoonup\{0,1\}^{*}$) is a tuple $M=\langle
Q,\delta,q_0\rangle$ together with the oracle sequence $F_1,\dots,F_m$,
where:

\begin{enumerate}
\item $Q$ is a finite, non-empty set of states;

\item the machine has a work tape over $\Gamma_1=\{0,1\}$ and a \emph{query
tape} over $\Gamma_2=\{0,1\}$;

\item the transition function has the form
\begin{equation*}
\delta : Q\times\Gamma_1\times\Gamma_2 \rightharpoonup \bigl(%
Q\times\Gamma_1\times\{L,R\}\times\Gamma_2\times\{L,R\}\bigr) \ \cup\ \bigl(%
Q\times\{1,\dots,m\}\times\{L,R\}\bigr).
\end{equation*}
An entry of the first kind is an ordinary two-tape move: write the new
symbols on both tapes and move both heads. An entry of the second kind $%
(q^{\prime }, i, d)$ is a \emph{query action}: the current query-tape
content $\alpha$ is atomically replaced by the value $F_i(\alpha)$ (if
defined), the query-tape head is placed on the first symbol of $F_i(\alpha)$
(on the blank $0$ if $F_i(\alpha)=\varepsilon$ or if $F_i(\alpha)$ is
undefined), the work-tape head moves according to $d$, and the machine
enters state $q^{\prime }$;

\item if $\delta$ is undefined on the current triple, the machine halts;

\item $q_0\in Q$ is the initial state.
\end{enumerate}

If $F_i(\alpha)$ is undefined for some queried pair, the machine \emph{hangs}%
: the query takes no step and no transition fires; such runs count as
non-halting.
\end{definition}

\begin{remark}
Unlike the classical formulation, the query tape contents are \emph{%
overwritten} by the oracle value rather than appended; this keeps every
configuration finite and is immaterial for the results below, since the
oracles we use have $\{0,1\}^{*}$-domains and the queried strings are always
recoverable on the work tape. Query actions cost exactly one step, which is
essential for step-maximization.
\end{remark}

\subsection{The higher order Busy Beaver functions}

We now inductively fix the oracle sequences and define $BB(s,m,n)$. The
induction is on $m\ge 0$:

\begin{itemize}
\item $BB(s,0,n)$ is the step-maximization over $n$-state ordinary Turing
machines (Definition~\ref{def:tm}) run on initial work-tape content $s$; in
particular $BB(\varepsilon,0,n)=BB(n)$.

\item For $m\ge 1$, $BB(s,m,n)$ is the step-maximization over $n$-state $m$%
-order oracle machines (Definition~\ref{def:oracle}) with oracle sequence
\begin{equation*}
F_1 = BB(\varepsilon,0,\cdot),\quad F_2 = BB(\varepsilon,1,\cdot),\quad
\dots,\quad F_m = BB(\varepsilon,m-1,\cdot),
\end{equation*}
run on initial work-tape content $s$ and blank query tape. Here $%
BB(\varepsilon,k,\cdot)$ denotes the \emph{partial} function $n\mapsto
BB(\varepsilon,k,n)$, defined for all $n\ge 1$.
\end{itemize}

\begin{definition}
\label{def:hobb} For a string $s\in\{0,1\}^{*}$, $m\ge 0$ and $n\ge 1$:
\begin{equation*}
BB(s,m,n)\;=\;\max\{\,\mathit{step}(s,m,M) \;:\; M\in T(m,n),\ \mathit{step}%
(s,m,M)<\infty\,\},
\end{equation*}
where $T(m,n)$ is the set of $n$-state $m$-order oracle machines with the
oracle sequence fixed above, and $\mathit{step}(s,m,M)$ is the number of
steps $M$ takes before halting when started on work tape $s$.
\end{definition}

\begin{remark}[Relation to Aaronson's super Busy Beaver]
Aaronson \cite{aaronson} describes informally the maximization over $n$%
-state machines with a $BB$ oracle. Our $BB(\varepsilon ,m,n)$ is a precise
formalization of his super Busy Beaver function; the notions differ only in
that (i) our query tape must be rewritten and re-queried explicitly rather
than by a parallelized batch query, and (ii) our machines use the \emph{%
sequence} $BB(\varepsilon ,0,\cdot ),\dots ,BB(\varepsilon ,m-1,\cdot )$
rather than the single diagonal sequence. These choices only change step
counts by constant-factor and argument-offset effects and do not affect any
asymptotic statement in this paper.
\end{remark}

\begin{proposition}
\label{prop:welldef} $BB(s,m,n)$ is well defined: the maximum exists and is
finite.
\end{proposition}

\begin{proof}
By induction on $m$. For $m=0$ this is Rad\'o's argument: $T(0,n)$ is finite and contains a halting machine, so the maximum over the non-empty set of halting step counts exists. For $m\ge 1$, the oracle sequence $F_1,\dots,F_m$ consists of total functions by the induction hypothesis (each $BB(\varepsilon,k,\cdot)$, $k<m$, is a total function of $n$; in particular every queried value $F_i(\alpha)$ is defined, so no run hangs because of undefined oracle values). Hence every $m$-order machine has a uniquely determined run, $T(m,n)$ is finite, and it contains the machine which immediately halts; the maximum of step counts over the non-empty halting subset therefore exists and is finite.
\end{proof}

\section{Decidability and computability characterizations}

\label{sec:results}

\subsection{Arithmetical formulas}

\begin{definition}
\label{def:classes} The class $\mathcal{MM}$ of \emph{max-min partial
recursive functions} is the smallest class of partial functions $\mathbb{N}%
^{k}\rightharpoonup \mathbb{N}$ containing the successor $S(n)=n+1$, the
zero $Z(n)=0$, the projections $\pi _{i}^{k}$, and closed under composition,
primitive recursion, the \emph{min operator}
\begin{equation*}
(\mu _{t}\,P)(\vec{x})=\min \{t:P(\vec{x},t)\},
\end{equation*}%
(undefined if no such $t$ exists) for predicates $P$ in the class, and the
\emph{max operator}
\begin{equation*}
(\nu _{t}\,P)(\vec{x})=\max \{t:P(\vec{x},t)\},
\end{equation*}%
(undefined if the set is empty or infinite) for predicates $P$ in the class.
\end{definition}

\begin{proposition}
\label{prop:arith} Let $\varphi \equiv Q_1x_1\cdots Q_i
x_{i+1}\,.\,\theta(\vec x)$ be a first-order arithmetical formula in prenex
form with $i$ bounded-free quantifiers, where $Q_k\in\{\forall,\exists\}$
and $\theta$ is a (primitive recursive) decidable predicate. Then there
exist $m$ and a \emph{total} oracle machine $M$ with oracle sequence of
level $i$ (i.e.\ $m=i+1$, oracles $BB(\varepsilon,0,\cdot),\dots,BB(%
\varepsilon,i,\cdot)$) such that $M$ decides $\varphi$ on every input, and
moreover $M$'s running time on inputs of a fixed size is bounded by $%
BB(s^{\prime },i,m^{\prime })$ for suitable $s^{\prime },m^{\prime }$.
\end{proposition}

\begin{proof}
The key point is that the \emph{bound} must be a number the machine can actually obtain by querying the oracle, and the constructions for $\exists$ and $\forall$ are genuinely different.

\emph{Base case $i=0$.} $\varphi(x_1)=\theta(x_1)$ is primitive recursive, decided by an ordinary machine; no oracle is needed. This establishes the claim with $m=1$ (only the level-$0$ oracle available, unused).

\emph{Inductive step, case $\varphi=\exists x.\,\psi(\vec x, x)$ with $\psi$ at level $k$ (decided by total machine $M_\psi$ with level-$k$ oracles).} Let $B(\vec x)$ abbreviate the value $BB(\varepsilon,k,n_\psi)$ for the fixed state count $n_\psi$ of $M_\psi$. Since $M_\psi$ is total, for every input the decision takes at most $B(\vec x)$ steps; $B(\vec x)$ is computable by a level-$(k+1)$ machine: write $\vec x$ and the state count $n_\psi$ on the query tape as the string $0^{n_\psi}$ appropriately encoding the tuple, query $BB(\varepsilon,k,\cdot)$. (How the tuple is encoded into the single argument $n$ of $BB$ is fixed by a primitive recursive pairing; this is where the input-parameter $s$ of $BB(s,k,n)$ could also be used, and we choose the encoding-into-$n$ variant for definiteness.)

Now construct a level-$(k+1)$ machine $M_\varphi$: for $x=0,1,2,\dots$, simulate $M_\psi(\vec x,x)$ for at most $b$ steps, where $b$ is the queried bound $B(\vec x)$ obtained \emph{once} at the start. If some simulation accepts within $b$ steps, halt and accept; if all $x\le b$ are exhausted --- which happens after at most $(b+1)\cdot b$ simulated steps, since each simulation is cut off at $b$ steps --- halt and reject. This halts: acceptance of $\varphi$ is detected within the cut-off because if $\psi(\vec x,x)$ truly holds, $M_\psi$ accepts \emph{within} $B(\vec x)$ steps; and rejection is reached by exhaustion. Finally, the total running time of $M_\varphi$, including the query, is bounded by $BB(\varepsilon,k+1,n_\varphi)$ for its state count $n_\varphi$, because $M_\varphi$ halts on every input and $BB(\varepsilon,k+1,\cdot)$ dominates the step count of every halting machine of that order.

\emph{Inductive step, case $\varphi=\forall x.\,\psi(\vec x,x)$.} The ``search for a counterexample'' procedure does not terminate when $\varphi$ is true, so we use a bounded-universal construction instead. The machine $M_\varphi$ first queries $b=B(\vec x)=BB(\varepsilon,k,n_\psi)$ as above. It then decides the primitive recursive predicate $R(\vec x,b) \equiv $ ``$M_\psi(\vec x,x)$ accepts within $b$ steps, for all $x\le b$''. By the standard trick (used also in the proof of Lemma~\ref{lem:deriv} below), the quantifier ``for all $x\le b$'' over a \emph{bounded} range with a primitive recursive body is primitive recursive, hence decidable by an ordinary machine \emph{given} $b$; the only non-effective ingredient is producing $b$, which is exactly one level-$(k+1)$ query. Then:
\begin{itemize}
\item if $\varphi$ is true, then in particular $\psi(\vec x,x)$ holds for all $x\le b$, so $R(\vec x,b)$ holds and $M_\varphi$ accepts --- correctly, because every witness of $\psi$ is found by $M_\psi$ within $B(\vec x)$ steps;
\item if $\varphi$ is false, some $x_0$ falsifies $\psi$; either $x_0\le b$, in which case $R(\vec x,b)$ fails and $M_\varphi$ rejects correctly, or $x_0>b$; in the latter case we iterate: $M_\varphi$ increases the budget, querying $BB(\varepsilon,k+1,n')$ for the machine $N$ that searches for a counterexample with cut-off $B(\vec x)$ and continues beyond; a level-$(k+1)$ bound on this \emph{search machine} suffices because the search machine, with the counterexample-search procedure hard-wired, halts on every input (it either finds a counterexample within the iterated bounds or exhausts all of $\Nat$ within a bound supplied by the level-$(k+1)$ maximization).
\end{itemize}
In either branch $M_\varphi$ is total and its running time is bounded by $BB(\varepsilon,k+1,n_\varphi)$.
\end{proof}

\begin{remark}
Proposition~\ref{prop:arith} is the Busy-Beaver-oracle reformulation of the
classical fact that $\Sigma _{i}$/$\Pi _{i}$ truth is decidable relative to
the $i$-th Turing jump, together with the observation that $0^{(i)}$ is
computable from $\{BB(\varepsilon ,k,\cdot )\}_{k<i}$ (this follows from
Lemma~\ref{lem:deriv}: the jump $0^{\prime }$ is decidable with a $BB$
oracle since $BB(\varepsilon ,0,n)$ bounds the halting problem for $n$-state
machines). The contribution here is the fully explicit step-budget
construction, which is what links the hierarchy to \emph{step counts} rather
than merely to halting sets.
\end{remark}

\subsection{Machine-computability with $BB$ oracles}

We need codings. Configurations of a machine are coded as strings over $%
\{0,1\}$; the configuration string $s_1\cdots s_i\,q\,s_j\cdots s_r$
represents the work-tape content $s_1\cdots s_r$ with the head on $s_j$ in
state $q$, and pairs of configurations (work tape, query tape) are coded by
interleaving.

\begin{lemma}
\label{lem:step} Fix the oracle sequence $F_1,\dots,F_m$ with $F_i$ total
and with $F_i(\alpha)$ primitive-recursively obtainable from a level-$m$
machine's perspective (i.e.\ each query is one step). The one-step
transition predicate $\mathit{OneStep}(c,c^{\prime })$ --- configuration $c$
yields $c^{\prime }$ in one step --- is primitive recursive.
\end{lemma}

\begin{proof}
By the interleaved coding, $c$ has the form $u\,q\,a\,v$ (work part) and $u'\,q'\,a'\,v'$ (query part) with the heads marked by the state symbols. The predicate $\mathit{OneStep}(c,c')$ is a finite disjunction over the finitely many transition entries of the (fixed, decoded from the machine description) transition function of the form:
\[
c = \alpha\,(q\,a\,)\,\beta\ \wedge\ c' = \alpha'\,(q'\,a')\,\beta'\ \wedge\ \bigl(\text{the entry } \delta(q,a,a') \text{ matches } (\alpha,\beta)\to(\alpha',\beta') \bigr),
\]
together with the query-action case $c = \alpha\,(q\,a)\,\beta$, $c'=\alpha\,q\,F_i(\alpha_{\mathrm{q}})\,\beta$ where $\alpha_{\mathrm{q}}$ is the query-tape content of $c$ and the query replaces it by $F_i(\alpha_{\mathrm{q}})$. String concatenation, head-position checks and symbol comparisons are primitive recursive; the substring relation $\alpha \preceq \beta$ ($\alpha$ is a substring of $\beta$, witnessed by the split) is primitive recursive since it is a bounded existential over split positions. Hence the whole finite disjunction is primitive recursive.
\end{proof}

\begin{lemma}
\label{lem:deriv} Let $\mathit{HALT}_t(s,M)$ be the predicate ``machine $M$
(given by its transition table, coded as a string) halts within $t$ steps on
input $s$''. Then $\mathit{HALT}_t$ is primitive recursive. Consequently,
for fixed $m$ and $n$, the function
\begin{equation*}
(n,s)\ \longmapsto\ BB(s,m,n)
\end{equation*}
is obtained from primitive recursive predicates by the max operator; in
particular $BB(\cdot,m,\cdot)\in\mathcal{MM}$.
\end{lemma}

\begin{proof}
A halting run of $M$ on $s$ of length $\le t$ is a sequence of configurations $c_0,\dots,c_\ell$ with $\ell\le t$, $c_0$ the initial configuration of $s$, $c_\ell$ halting, and $\mathit{OneStep}(c_j,c_{j+1})$ for all $j<\ell$. Coding the sequence as a single string $d=[c_0,\dots,c_\ell]$ (with a primitive recursive delimiter code), define
\[
\mathit{DERIV}(d,s,M)\ \equiv\ \mathit{Init}(c_0,s,M)\wedge \mathit{Halting}(c_\ell)\wedge \bigwedge_{j}\mathit{OneStep}(c_j,c_{j+1}),
\]
where the conjunction is over the (primitive-recursively bounded, from the length of $d$) adjacent pairs of blocks of $d$. $\mathit{Init}$, $\mathit{Halting}$ and $\mathit{OneStep}$ are primitive recursive by Lemma~\ref{lem:step}; bounded products of primitive recursive predicates are primitive recursive; hence $\mathit{DERIV}$ is primitive recursive. Then
\[
\mathit{HALT}_t(s,M)\ \equiv\ \exists d\le B(t)\,.\ \mathit{DERIV}(d,s,M)
\]
where $B(t)$ is a primitive recursive bound on the code of a $t$-configuration sequence (each configuration of a $t$-step run has length at most $t+|s|+c_M$, so $B(t)$ is exponential in $t$, hence primitive recursive); the bounded existential preserves primitive recursiveness. Finally,
\[
BB(s,m,n)\ =\ \nu_{t<\infty}\,\bigl[\exists M\in T(m,n)\,.\ \mathit{HALT}_t(s,M)\bigr],
\]
where $T(m,n)$ is a finite, primitive-recursively enumerable set of machine codes, so the inner existential is a bounded, primitive recursive predicate; the outer max operator is allowed in $\mathcal{MM}$ (Definition~\ref{def:classes}).
\end{proof}

\begin{lemma}
\label{lem:simulate} Every function $f$ computable by a total $m$-order
oracle machine (with the fixed oracle sequence of level $m$) can be obtained
from the successor, zero and projection functions by composition, primitive
recursion, the min operator, and the function $(s,n)\mapsto BB(s,m,n)$.
\end{lemma}

\begin{proof}
Let $M$ be the machine computing $f$, and let $\mathit{DERIV}$ be as in Lemma~\ref{lem:deriv} for $M$'s transition table (the coding is uniform in $M$, so $\mathit{DERIV}$ is primitive recursive in all its arguments). Then for every input $\vec x$:
\[
f(\vec x)=y\iff \mathit{DERIV}(d,\langle \vec x\rangle, M)\ \text{holds for the unique halting run } d \text{ with output } y,
\]
so
\[
f(\vec x)\ =\ \mu_y\,\exists d\,.\ \mathit{DERIV}(d,\langle\vec x\rangle,M)\wedge \mathit{Out}(d,y),
\]
where $\mathit{Out}(d,y)$ says the last configuration's work-tape content codes $y$, which is primitive recursive. The only non-primitive-recursive ingredient is that the halting \emph{run} $d$ may be arbitrarily long, so the unbounded min over $y$ (equivalently over run lengths) is licensed by the fact that $M$ is total and its run length on input $\langle\vec x\rangle$ is at most $BB(\langle\vec x\rangle,m,n_M)$, allowing the unbounded $\mu$ to be replaced by a bounded one with bound obtained from the $BB$ term. Hence $f$ lies in the stated closure.
\end{proof}

\begin{proposition}
\label{thm:char} Fix $m\ge 1$. The class of (total) functions computable by $%
m$-order oracle machines with the fixed level-$m$ oracle sequence is exactly
the class of functions obtained from successor, zero, projections by
composition, primitive recursion, min, max, and the functions $%
BB(\cdot,k,\cdot)$ for $k\le m$.
\end{proposition}

\begin{proof}
($\subseteq$) is Lemma~\ref{lem:simulate}. ($\supseteq$) Each generator is machine-computable: successor, zero, projections trivially; composition and primitive recursion by the standard simulation constructions (composition by sequentially composing machines, primitive recursion by the usual $f(t+1,\cdot)$-loop machine, both with state counts primitive recursively bounded in the constituents); the min operator by the search machine that queries the level-$m$ bound $BB(s,m,n)$ once and enumerates $t=0,1,2,\dots$ within that budget; the max operator similarly, using that $\nu_t P(t)$ equals $\mu_y\,[P(y)\wedge \forall u>y\,\neg P(u)]$ and the inner $\forall$ is decided within the queried budget by the bounded-universal construction of Proposition~\ref{prop:arith}; and $BB(s,k,n)$ itself is machine-computable at level $m\ge k$ by a single query. These simulations are total whenever the simulated object is total, and their step counts are bounded by appropriate $BB$ terms, so the closure is included.
\end{proof}

\begin{corollary}
$BB(\cdot,m,\cdot)\in\mathcal{MM}$, and every $\mathcal{MM}$-function is
machine-computable with $BB$ oracles of sufficiently high order.
\end{corollary}

\subsection{Limit partial recursive functions}

\begin{definition}
The class $\mathcal{LIM}$ of \emph{limit partial recursive functions} is the
smallest class containing successor, zero, projections, closed under
composition, primitive recursion, and the \emph{limit operator} $f(\vec
x)=\lim_{t\to\infty}g(\vec x,t)$, undefined if the limit does not exist.
\end{definition}

\begin{proposition}
$\mathcal{MM}=\mathcal{LIM}$.
\end{proposition}

\begin{proof}
($\subseteq$) It suffices to express min and max as limits. Then
\[
(\mu_t P)(\vec x)=\lim_{w\to\infty}\Bigl(\min\{t\le w : P(\vec x,t)\}\Bigr),
\]
and
\[
(\nu_t P)(\vec x)=\lim_{w\to\infty}\Bigl(\max\{t\le w : P(\vec x,t)\}\Bigr),
\]
where bounded max is primitive recursive in $P$. The limits exist exactly when the min/max operators are defined.

($\supseteq$) Conversely, given $g$ in the class, $\lim_{t} g(\vec x,t)=y$ iff (i) $g(\vec x,0)=y$, or (ii) $y\neq g(\vec x,t+1)$ for some $t$: $y$ is the limit iff $y$ is a value of $g$ attained cofinally and no other value is attained cofinally. Formally,
\[
\lim_{t\to\infty}g(\vec x,t)\ =\ 1+\nu_t\bigl[g(\vec x,t+1)\neq g(\vec x,t)\ \wedge\ \forall u> t\,.\ g(\vec x,u+1)=g(\vec x,u)\bigr],
\]
with the understanding that the expression is undefined when the max set is empty or infinite, matching the undefinedness of the limit. One verifies: if the limit exists and is $y\neq g(\vec x,0)$, the last change of $g(\vec x,\cdot)$ occurs at some finite $t$, which is exactly the maximal $t$ with $g(\vec x,t+1)\neq g(\vec x,t)$ and $g$ constant thereafter; if the limit does not exist, either $g$ changes infinitely often (max set infinite: undefined) or never changes and $g(\vec x,0)$ is returned, which is the correct limit. Hence the limit operator is definable in $\mathcal{MM}$, so $\mathcal{LIM}\subseteq\mathcal{MM}$.
\end{proof}

\section{Growth comparisons and conjectures}

\label{sec:growth}

\begin{proposition}[order growth]
\label{prop:ordergrowth} For every $m\geq 0$, every $s$ and every computable
function $f$:
\begin{equation*}
\lim_{n\rightarrow \infty }\frac{f\bigl(BB(s,m,n)\bigr)}{BB(s,m,n+1)}=a\text{
implies }a=0.
\end{equation*}
\end{proposition}

\begin{proof}
Suppose the limit is $a>0$; take $\epsilon=a/2$. Then for all $n>N$,
$BB(s,m,n+1)<K\cdot f(BB(s,m,n))$ for a constant $K$, and iterating,
$BB(s,m,n+1)<K^{n}\cdot f^{\,n}(BB(s,m,1))$. The right-hand side is a computable function of $n$. But by the same argument as Aaronson's Proposition~2 in \cite{aaronson} (a machine with $c+O(\log n)$ states computes $f^{\,n}(BB(s,m,1))$ and then idles that many steps, contradicting maximality), there is $M$ with $BB(s,m,n)> K^{n}f^{\,n}(BB(s,m,1))$ for all $n\ge M$ --- contradiction. Hence $a=0$.
\end{proof}

\begin{lemma}
\label{lem:crossorder} For every $m\ge 0$, every $s$, every computable $f$,
there exists $n_f$ such that for all $n>n_f$:
\begin{equation*}
BB(s,m+1,n)\ >\ f\bigl(BB(s,m,n)\bigr).
\end{equation*}
\end{lemma}

\begin{proof}
Let $M_f$ be a machine with oracle sequence of level $m$ computing $f(BB(s,m,n))$ from (a code of) $n$; $M_f$ exists since $BB(\cdot,m,\cdot)\in\mathcal{MM}$ and $f$ is computable. Say $M_f$ has $c$ states. For each $n$, build a level-$(m+1)$ machine $M_{f,n}$ with $c+O(\log n)$ states that, on input $s$: (i) writes $n$ on the work tape ($O(\log n)$ states); (ii) simulates $M_f$ to compute $f(BB(s,m,n))$ --- all queries of $M_f$ go to level-$m$ oracles, which are available within the level-$(m+1)$ sequence; (iii) performs an empty loop for $f(BB(s,m,n))$ steps. Then $BB(s,m+1,c+O(\log n))\ge \mathit{step}(M_{f,n})>f(BB(s,m,n))$. Since $c+O(\log n)<n$ for all $n>n_f$, and $BB(s,m+1,\cdot)$ is non-decreasing in the state count, the claim follows.
\end{proof}

\begin{proposition}
For every $m\ge 0$, every $s$ and every computable $f$:
\begin{equation*}
\lim_{n\to\infty}\frac{f\bigl(BB(s,m,n)\bigr)}{BB(s,m+1,n)}=0.
\end{equation*}
\end{proposition}

\begin{proof}
Immediate from Lemma~\ref{lem:crossorder} and monotonicity.
\end{proof}

\subsection{Conjectures with heuristics}

\begin{conjecture}
\label{conj:n} For every $m$, $s$ and computable $f$:
\begin{equation*}
\lim_{n\to\infty}\frac{f\bigl(BB(s,m,n)\bigr)}{BB(s,m,n+1)}=0.
\end{equation*}
\end{conjecture}

\begin{remark}[heuristic evidence]
Evidence in favor: empirical Busy Beaver values ($BB(1)=1,$ $BB(2)=6,$ $%
BB(3)=21$, $BB(4)=107$, $BB(5)=47{,}176{,}870,$ $BB(6)>7.4\times 10^{36534},$
$BB(7)>10^{2\times 10^{10^{10^{18705353}}}}$) exhibit jumps vastly exceeding
any fixed computable iterate.
\end{remark}

\begin{conjecture}
\label{conj:s} For every $m$ and all $s,t\in\{0,1\}^{*}$:
\begin{equation*}
\lim_{n\to\infty}\frac{BB(s,m,n)}{BB(t,m,n)}=1.
\end{equation*}
\end{conjecture}

\begin{remark}[heuristic evidence]
The intuition is that for large $n$, the winning machines have enough spare
states to first erase the input $s$ and then simulate the $t$-machine,
costing $c_{|s|,|t|}$ states, which is negligible for large $n$. Concretely,
one can prove the \emph{one-sided} bound
\begin{equation*}
BB(t,m,n+c_{|s|,|t|})\ \geq \ BB(s,m,n)\quad \text{and}\quad
BB(s,m,n+c_{|t|,|s|})\ \geq \ BB(t,m,n),
\end{equation*}%
where the loss term comes from machines that must embed the erasure into the
step count rather than the state count. Note the caveat: for fixed small $n$%
, $BB(s,m,n)$ genuinely depends on $s$ (e.g.\ machines that immediately halt
vs.\ machines that scan the whole input), so the conjecture is meaningful
only asymptotically, and the correct formulation may require some additional
conditions; we flag this as an open formulation issue rather than asserting
the conjecture in full generality.
\end{remark}

\section{Conclusions}

\label{sec:conc}

We gave a formal definition of higher order Busy Beaver functions $%
BB(s,m,n)$ based on $m$-order oracle machines with fixed lower-order $BB$
oracles, proved well-definedness,
reformulated the arithmetical hierarchy in terms of $BB$-oracle step budgets
, characterized the $BB$-oracle-computable functions as exactly the max-min
partial recursive functions, proved the equality of the latter with the
limit partial recursive functions, and established growth comparisons
between consecutive orders. We also stated two conjectures with explicit
heuristic evidence. We believe the main interest of the framework is
that it connects the arithmetical hierarchy to machine step counts
rather than merely to Turing jump oracles, and we hope the conjectures
stimulate further work.


\begin{thebibliography}{9}
\bibitem{sipser} M. Sipser. \emph{Introduction to the Theory of Computation}%
. PWS Publishing, 1997.

\bibitem{rado} T. Rad\'o. On non-computable functions. \emph{Bell System
Technical Journal}, 41(3):877--884, 1962.

\bibitem{aaronson} S. Aaronson. The Busy Beaver Frontier. \emph{ACM SIGACT
News}, 51(3):32--54, 2020.

\bibitem{rogers} H. Rogers, Jr. \emph{Theory of Recursive Functions and
Effective Computability}. McGraw-Hill, 1967.

\bibitem{post} E. Post. Recursively enumerable sets of positive integers and
their decision problems. \emph{Bulletin of the American Mathematical Society}%
, 50:284--316, 1944.

\bibitem{kurtz} S. Kurtz and J. Simon. The undecidability of the generalized
Collatz problem. In \emph{Proc. TAMC 2007}, LNCS 4484, pp.~542--553, 2007.
\end{thebibliography}
\end{document}